\documentclass[12pt]{article}
\usepackage{graphicx}

\begin{document}

\title{An entangling-probe attack on Shor's algorithm for factorization}

\author{Hiroo Azuma\thanks{Email: hiroo.azuma@m3.dion.ne.jp}
\\
\\
{\small Advanced Algorithm \& Systems Co., Ltd.,}\\
{\small 7F Ebisu-IS Building, 1-13-6 Ebisu, Shibuya-ku, Tokyo 150-0013, Japan}\\
}

\date{\today}

\maketitle

\begin{abstract}
We investigate how to attack Shor's quantum algorithm for factorization
with an entangling probe.
We show that an attacker can steal
an exact solution of Shor's algorithm outside an institute where the quantum computer is installed
if he replaces its initialized quantum register with entangled qubits,
namely the entangling probe.
He can apply arbitrary local operations
to his own probe.
Moreover,
we assume that there is an unauthorized person who helps the attacker to commit a crime inside the institute.
He tells garbage data obtained from measurements of the quantum register to the attacker secretly behind a legitimate user's back.
If the attacker succeeds in cracking Shor's algorithm,
the legitimate user obtains a random answer and does not notice the attacker's illegal acts.
We discuss how to detect the attacker.
Finally we estimate a probability that the quantum algorithm inevitably makes an error,
of which the attacker can take advantage.
\end{abstract}

\section{\label{section-introduction}Introduction}
Utilization of entanglement to steal information of the quantum cryptographic protocol in key distribution
has been investigated in several strategies
\cite{Fuchs1996,Slutsky1998,Brandt2005,Shapiro2006,Kim2007}.
In these attacks,
first,
Eve (an eavesdropper) lets a carrier qubit of Alic and Bob (a sender and a receiver) interact with her own qubit,
called as a probe,
so that the carrier qubit and the probe are in an entangled state.
Next,
Eve measures her own probe and obtains information about the key distribution.
In Ref.~\cite{Fuchs1996},
the general eavesdropping attack with the probe and its trade-off between the information gain and disturbance are investigated.
In Ref.~\cite{Slutsky1998},
individual attacks with the probe
against the four-state Bennett-Brassard 1984 (BB84) and the two-state Bennett 1992 (B92) quantum cryptographic protocols are examined
\cite{Bennett1984,Bennett1992}.
In Ref.~\cite{Brandt2005},
for the interaction of the entangling probe with signal basis states of the BB84 protocol,
some optimized unitary transformations are calculated.
In Ref.~\cite{Shapiro2006},
it is discussed how to realize the Fuchs-Peres-Brandt (FPB) entangling probe physically
using a deterministic controlled-NOT gate implemented with single-photon two-qubit (SPTQ) quantum logic.
In Ref.~\cite{Kim2007},
physical simulation of the FPB probe with the SPTQ logic is discussed,
including physical errors.
Moreover,
as an advanced version,
Eve can let her probe interact with more than one of Alice and Bob's qubits simultaneously,
which is called as a coherent attack
\cite{Cirac1997}.
In Ref.~\cite{Hwang2001},
Eve's optimal information gained in coherent attacks is estimated for the six-state protocol,
which is another version of the BB84 protocol and discussed in Ref.~\cite{Bruss1998}.
It is considered that coherent measurements can supply much more information than incoherent individual measurements can do
\cite{Massar1995}.
Because Eve can steal information from quantum cryptographic protocols with entangling probes as mentioned above,
we may expect to do exactly same for attacking quantum algorithms.
This is the primary motivation of the current paper.

In this paper,
we examine what kind of security incidents arise if the attacker replaces initialized qubits of the quantum register of the quantum computer with entangled qubits,
that is to say the entangling probe,
for Shor's factoring algorithm
\cite{Shor1997,Ekert1996}.
We formulate circumstances where the attacker commits illegal acts as follows.
We assume that the quantum computer is installed in a research institute and connected to external quantum networks located outside the institute.
We assume that the attacker can replace the qubits in the initial state belonging to the quantum register of the quantum computer with the entangled ones
via the networks.
Thus, the attacker shares entanglement between the qubits of the quantum register and the qubits that he holds.
The attacker can apply local operations,
for example,
arbitrary unitary transformations and measurements,
to his own qubits.
Moreover,
we assume that there is an unauthorized person who helps the attacker to commit illegal acts harming the quantum algorithm inside the institute.
The accomplice tells garbage data obtained as results of measurements of the quantum register to the attacker outside the institute secretly
behind a legitimate user's back.

We assume that the attacker wants to achieve the following two goals.
The first one is to steal an exact solution of the quantum algorithm that the legitimate user performs.
The second one is to let the legitimate user not notice the attacker's illegal acts.
In addition,
we attach a condition to the attacker's plan of the malicious acts.
During the sequence of the illegal acts,
the accomplice does not obtain any information about the exact solution of the quantum algorithm.

First of all,
we discuss how to attack Simon's algorithm
\cite{Simon1997,Aharonov1999}.
Next,
we consider how to attack Shor's factoring algorithm.
This is because Shor's algorithm is based on Simon's algorithm and the quantum discrete Fourier transform
\cite{Coppersmith1994}.
Furthermore,
we discuss how to detect the attacker's illegal acts.
Finally,
we both numerically and analytically estimate a probability that Shor's algorithm inevitably makes an error,
of which the attacker can take advantage.
The probability that Shor's algorithm fails is approximately equal to $0.2263$.

This paper is organized as follows.
In Sec.~\ref{section-review-Simon},
we give a brief review of Simon's algorithm.
In Sec.~\ref{section-review-Shor},
we give a brief review of Shor's algorithm for factorization.
In Sec.~\ref{section-attack-Simon},
we consider how to attack Simon's algorithm.
In Sec.~\ref{section-attack-Shor},
we consider how to attack Shor's algorithm for factorization.
In Sec.~\ref{numerical-calculations},
we evaluate a probability that Shor's algorithm outputs a wrong answer.
The attacker makes use of this chance for committing a crime.
In Sec.~\ref{section-discussion},
we give a brief discussion.

\section{\label{section-review-Simon}A brief review of Simon's algorithm}
In this section,
we give a brief review of Simon's algorithm
\cite{Simon1997,Aharonov1999}.
It solves the following problem efficiently.
We consider an unknown Boolean function
$f:\{0,1\}^{n}\to\{0,1\}^{n}$
for some integer $n$.
We assume that the function $f$ satisfies one of the next two conditions.
\begin{itemize}
\item
All $\{f(x):x\in\{0,1\}^{n}\}$ are different and the function $f$ is one-to-one.
\item
There exists a certain nontrivial value $s\in\{0,1\}^{n}$ where $s\neq 0$,
and the statement
``$f(x)=f(y)$ $\Leftrightarrow$ $x=y$ or $x=y\oplus s$"
holds true.
Thus,
the function $f$ is two-to-one.
\end{itemize}
We want to distinguish between the above two cases with the fewest evaluations of $f$ possible.
If the latter case holds,
we also want to obtain $s$.

Simon's algorithm
is given as follows.
We prepare two $n$-qubit registers A and B.
As an operation of evaluating the function $f$,
we introduce a unitary gate $U_{f}$,
\begin{equation}
|x\rangle_{\mbox{\scriptsize A}}
|y\rangle_{\mbox{\scriptsize B}}
\stackrel{U_{f}}{\longrightarrow}
|x\rangle_{\mbox{\scriptsize A}}
|y\oplus f(x)\rangle_{\mbox{\scriptsize B}},
\label{U_f-gate_01}
\end{equation}
where
$x,y\in\{0,1\}^{n}$.
We define the Hadamard transformation for one qubit as
\begin{equation}
H
=
\frac{1}{\sqrt{2}}
\left(
\begin{array}{cc}
1 & 1 \\
1 & -1 \\
\end{array}
\right),
\end{equation}
where we use an orthonormal basis
$\{|0\rangle,|1\rangle\}$
for the matrix representation.
Moreover,
we introduce the following notations.
$\forall x,y\in\{0,1\}^{n}$,
we define
$x\oplus y=(
x_{1}+y_{1} \bmod 2,
x_{2}+y_{2} \bmod 2,
...,
x_{n}+y_{n} \bmod 2)$
and
$x\cdot y=\sum_{i=1}^{n}x_{i}y_{i} \bmod 2$.

First,
we initialize the quantum registers A and B in the following state:
\begin{equation}
\frac{1}{\sqrt{2^{n}}}
\sum_{x\in\{0,1\}^{n}}
|x\rangle_{\mbox{\scriptsize A}}
|0\rangle_{\mbox{\scriptsize B}}.
\label{initial-state-Simon-0}
\end{equation}
Second,
we apply the unitary gate $U_{f}$ defined in Eq.~(\ref{U_f-gate_01}) to the above state and obtain
\begin{equation}
\stackrel{U_{f}}{\longrightarrow}
\frac{1}{\sqrt{2^{n}}}
\sum_{x\in\{0,1\}^{n}}
|x\rangle_{\mbox{\scriptsize A}}
|f(x)\rangle_{\mbox{\scriptsize B}}.
\end{equation}
Third,
we apply the Hadamard transformation to each qubit of the quantum register A,
\begin{equation}
\stackrel{H^{\otimes n}}{\longrightarrow}
\frac{1}{2^{n}}
\sum_{x,y\in\{0,1\}^{n}}
(-1)^{x\cdot y}
|y\rangle_{\mbox{\scriptsize A}}
|f(x)\rangle_{\mbox{\scriptsize B}}.
\end{equation}
Fourth,
we observe the quantum register A and obtain $y=k_{1}$.
Repeating similar operations $m$ times,
we obtain $k_{1},k_{2}, ..., k_{m}$,
where $m$ is of order $O(n)$.

We consider a finite set of equations,
\begin{eqnarray}
k_{1}\cdot s
&=&
0 \bmod 2, \nonumber \\
k_{2}\cdot s
&=&
0 \bmod 2, \nonumber \\
&\vdots& \nonumber \\
k_{m}\cdot s
&=&
0 \bmod 2.
\label{equations-for-s}
\end{eqnarray}
If we find nontrivial $s(\neq 0)$ that satisfies the above system of equations,
the function $f$ is two-to-one.
Otherwise,
for $s=0$,
it is one-to-one.

The reason why the facts mentioned above hold is as follows.
If the function $f$ is one-to-one,
$\mbox{Prob}(k)$,
a probability that we obtain $k_{1}=k$,
is given by
\begin{equation}
\mbox{Prob}(k)
=
\frac{1}{2^{n}}
\quad
\forall k\in\{0,1\}^{n}.
\end{equation}
Contrastingly,
if the function $f$ is two-to-one,
it is given by
\begin{eqnarray}
\mbox{Prob}(k)
&=&
\frac{1}{2}
\sum_{x\in\{0,1\}^{n}}
\frac{1}{2^{2n}}
|
(-1)^{x\cdot k}
+
(-1)^{(x\oplus s)\cdot k}
|^{2} \nonumber \\
&=&
\left\{
\begin{array}{lll}
1/2^{n-1} & \quad & k\cdot s=0 \bmod 2, \\
0         & \quad & \mbox{otherwise.} \\
\end{array}
\right.
\end{eqnarray}
Thus,
we obtain Eq.~(\ref{equations-for-s}).

\section{\label{section-review-Shor}A brief review of Shor's algorithm for factorization}
In this section,
we give a brief review of Shor's algorithm for factorization
\cite{Shor1997,Ekert1996}.
We let $N$ is an odd composite integer.
Throughout this paper, we assume $N\gg 1$.
To factorize large $N$ efficiently,
Shor's algorithm solves the following equivalent problem.
We choose an integer $y$ at random,
where $y$ is coprime to $N$
[$\gcd(y,N)=1$]
and $2\leq y\leq N-1$.
Then,
we define a function,
\begin{equation}
F_{N}(a)\equiv y^{a} \bmod N.
\label{y-a-mod-N}
\end{equation}
Shor's algorithm finds a period $r$ of $F_{N}(a)$
with polynomial time in $\log N$.
If $r$ is even and
$y^{r/2}\not\equiv\pm 1 \bmod N$,
$\gcd(y^{r/2}\pm 1,N)$
is a nontrivial factor of $N$.
A probability that $y$ is odd or
$y^{r/2}\equiv\pm 1 \bmod N$
is low enough for efficient computation.

We can regard Shor's algorithm as an combination of Simon's algorithm and the discrete Fourier transform.
To obtain the period $r$ of $F_{N}(a)$ efficiently,
we carry out a process described below.
First, we choose an integer $q=2^{L}$ as a power of two,
where $N^{2}\leq q\leq 2N^{2}$.
Then,
we initialize the quantum registers A and B in the following state:
\begin{equation}
\frac{1}{\sqrt{q}}
\sum_{a=0}^{q-1}
|a\rangle_{\mbox{\scriptsize A}}
|0\rangle_{\mbox{\scriptsize B}}.
\label{Shor-1st-step}
\end{equation}
Second,
we compute $F_{N}(a)$ given by Eq.~(\ref{y-a-mod-N}) and write down its output on the quantum register B,
\begin{equation}
\frac{1}{\sqrt{q}}
\sum_{a=0}^{q-1}
|a\rangle_{\mbox{\scriptsize A}}
|y^{a} \bmod N\rangle_{\mbox{\scriptsize B}}.
\label{Shor-2nd-step}
\end{equation}

Third,
we observe the quantum register B.
We suppose that we obtain
$z\equiv y^{l} \bmod N$ for some least $l$.
Then,
the quantum register A irreversibly reduces to the following state:
\begin{equation}
|\phi_{l}\rangle_{\mbox{\scriptsize A}}
=
\frac{1}{\sqrt{A+1}}
\sum_{j=0}^{A}
|jr+l\rangle_{\mbox{\scriptsize A}},
\label{third-steo-Shor-algorithm-01}
\end{equation}
where $A$ is given by
\begin{equation}
A=\lfloor (q-l)/r\rfloor.
\label{definition-A}
\end{equation}
We rewrite Eq.~(\ref{third-steo-Shor-algorithm-01}) in the form,
\begin{equation}
|\phi_{l}\rangle_{\mbox{\scriptsize A}}
=
\sum_{a=0}^{q-1}
f(a)|a\rangle_{\mbox{\scriptsize A}},
\label{third-steo-Shor-algorithm-02}
\end{equation}
\begin{equation}
f(a)
=
\left\{
\begin{array}{ll}
1/\sqrt{A+1} & \mbox{for $a=jr+l$, $j=0,1,...,A$,} \\
0 & \mbox{otherwise.}
\end{array}
\right.
\label{definition-f(a)}
\end{equation}

Fourth,
we apply the discrete Fourier transform (DFT) to the quantum register A given by Eq.~(\ref{third-steo-Shor-algorithm-02}) and obtain
\begin{equation}
\mbox{DFT}_{q}|\phi_{l}\rangle_{\mbox{\scriptsize A}}
=
\sum_{c=0}^{q-1}
\tilde{f}(c)|c\rangle_{\mbox{\scriptsize A}},
\label{DFT-state-Shor}
\end{equation}
\begin{equation}
\tilde{f}(c)
=
\frac{1}{\sqrt{q}}
\sum_{a=0}^{q-1}
\exp(2\pi i ac/q)f(a).
\label{Shor-amplitude}
\end{equation}
Fifth,
we observe the quantum register A.
We suppose that we obtain $c$ with the measurement.
We consider values of $c$ which satisfy the following inequality:
\begin{equation}
-\frac{r}{2}
\leq
rc \bmod q
\leq
\frac{r}{2}.
\label{range_rc_mod_q}
\end{equation}
The number of values of $c \bmod q$ that satisfy Eq.~(\ref{range_rc_mod_q}) is $r$.
We can observe each value of $c$ that satisfies Eq.~(\ref{range_rc_mod_q}) at least with a probability of $(4/\pi^{2})(1/r)$.
Thus,
we can observe $c$ that satisfies Eq.~(\ref{range_rc_mod_q}) at least with a total probability of $4/\pi^{2}\simeq 0.405$.
To put it the other way around,
we can observe $c$ that does not satisfy Eq.~(\ref{range_rc_mod_q}) at most with a probability of $1-(4/\pi^{2})\simeq 0.595$.

In the simplified situation where $r$ divides $q$ exactly,
we can observe $c$ satisfying Eq.~(\ref{range_rc_mod_q}) with a zero error probability.
(If $r$ divides $q$ exactly,
there are $r$ values of $c$ that satisfy $rc \bmod q=0$.)
Otherwise,
the operation mentioned above certainly makes an error,
that is to say supplies $c$ not satisfying Eq.~(\ref{range_rc_mod_q}),
with a non-zero probability.
In Sec.~\ref{numerical-calculations},
we evaluate this probability both numerically and analytically.

If Eq.~(\ref{range_rc_mod_q}) holds,
there is only one value of $c'$ that satisfies
\begin{equation}
|rc-c'q|\leq\frac{r}{2},
\label{condition-rc}
\end{equation}
where $0\leq c'\leq r-1$.
Then,
$c'/r$ is a convergent of the continued fraction of $c/q$.
If $\gcd(c',r)=1$,
that is to say $c'$ is coprime to $r$,
we obtain $r$.
The number of values of $c'$ that satisfy $\gcd(c',r)=1$ is given by $\phi(r)$, Euler's Phi function.
Thus,
a probability that we choose $c'$ satisfying $\gcd(c',r)=1$ is given by $\phi(r)/r$.
According to Ref.~\cite{Hardy1975},
the average order of $\phi(r)$ is given by $6r/\pi^{2}$,
and we obtain
$\phi(r)/r\simeq 6/\pi^{2}\simeq 0.608$.
Or, in other words,
we cannot obtain $r$ with a probability of $0.392$ around.

\section{\label{section-attack-Simon}How to attack Simon's algorithm}
In this section,
we discuss how to attack Simon's algorithm.
First of all,
we assume that the number of qubits in the quantum register A,
namely $n$,
is publicly disclosed,
so that the attacker knows it.
We consider the case where the attacker replaces the qubits of the register A
with the entangled ones that keep entanglement between quantum registers A and C.
We assume that the quantum register C is an $n$-qubit system the attacker holds.
The attacker prepares the following state instead of Eq.~(\ref{initial-state-Simon-0}):
\begin{equation}
\frac{1}{\sqrt{2^{n}}}
\sum_{x\in\{0,1\}^{n}}
|x\rangle_{\mbox{\scriptsize A}}
|x\rangle_{\mbox{\scriptsize C}}
|0\rangle_{\mbox{\scriptsize B}}.
\label{initial-state-Simon-attacker-0}
\end{equation}
To generate the above state,
the attacker has only to replace the $i$th qubit of the register A with the maximally entangled state,
\begin{equation}
|\Phi^{(+)}\rangle_{\mbox{\scriptsize A}i\mbox{\scriptsize C}i}
=
(1/\sqrt{2})
(|0\rangle_{\mbox{\scriptsize A}i}|0\rangle_{\mbox{\scriptsize C}i}
+
|1\rangle_{\mbox{\scriptsize A}i}|1\rangle_{\mbox{\scriptsize C}i})
\quad
\mbox{for $i=1,...,n$}.
\end{equation}

The legitimate user applies the unitary gate $U_{f}$ defined in Eq.~(\ref{U_f-gate_01}) to the quantum registers A and B
in Eq.~(\ref{initial-state-Simon-attacker-0})
and we obtain
\begin{equation}
\stackrel{U_{f}}{\longrightarrow}
\frac{1}{\sqrt{2^{n}}}
\sum_{x\in\{0,1\}^{n}}
|x\rangle_{\mbox{\scriptsize A}}
|x\rangle_{\mbox{\scriptsize C}}
|f(x)\rangle_{\mbox{\scriptsize B}}.
\end{equation}
Next,
the legitimate user and the attacker apply the Hadamard transformation to each qubit of the quantum registers A and C, respectively and independently.
Then,
we obtain the state of the whole system as
\begin{equation}
\stackrel{H_{\mbox{\tiny A}}^{\otimes n}\otimes H_{\mbox{\tiny C}}^{\otimes n}}{\longrightarrow}
\frac{1}{2^{3n/2}}
\sum_{x,y,z\in\{0,1\}^{n}}
(-1)^{x\cdot(y\oplus z)}
|y\rangle_{\mbox{\scriptsize A}}
|z\rangle_{\mbox{\scriptsize C}}
|f(x)\rangle_{\mbox{\scriptsize B}}.
\end{equation}
We suppose that results of the measurements of the quantum registers A and C are given by $y=k_{1}$ and $z=j_{1}$,
respectively.
We repeat similar operations for $m$ times and obtain
$k_{1},k_{2}, ..., k_{m}$
and
$j_{1},j_{2}, ..., j_{m}$.

Here,
we think about the following finite set of equations:
\begin{eqnarray}
(k_{1}\oplus j_{1})\cdot s
&=&
0 \bmod 2, \nonumber \\
(k_{2}\oplus j_{2})\cdot s
&=&
0 \bmod 2, \nonumber \\
&\vdots& \nonumber \\
(k_{m}\oplus j_{m})\cdot s
&=&
0 \bmod 2.
\label{equations-for-s-attacker}
\end{eqnarray}
If we find nontrivial $s(\neq 0)$ that satisfies the above system of equations,
the function $f$ is two-to-one.
Otherwise, for $s=0$, it is one-to-one.

We consider the following scenario for attacking the quantum computer.
Before the legitimate user performs Simon's algorithm,
the attacker replaces the qubits of the quantum register A with the entangled ones as mentioned above.
The legitimate user obtains random integers $\{k_{1},...,k_{m}\}$.

Under these circumstances,
the legitimate user cannot detect the attacker's illegal acts.
The reason is as follows.
Because the legitimate user obtains the random integers $\{k_{1},...,k_{m}\}$,
the solution of Eq.~(\ref{equations-for-s}) is given by $s=0$.
In other words,
if the legitimate user observes plenty of the random integers $\{k_{1},...,k_{m}\}$,
they force him to obtain $s=0$ as the solution of Eq.~(\ref{equations-for-s}).
Thus,
he has to accept a conclusion that $f$ is one-to-one.
Even for such a large $m$,
$m\sim O(n)$ holds.

If the legitimate user once obtains $s=0$,
he cannot confirm
whether or not it is a correct answer practically.
This is because he has to find two different numbers $x$ and $x\oplus s$ such that $f(x)=f(x\oplus s)$ for $s\neq 0$
among $2^{n}$ integers for proving that $s=0$ is wrong.
$\forall x\in\{0,1\}^{n}$,
there exists at most only one $y(\neq x)$
that satisfies $f(x)=f(y)$.
Thus,
it is practically impossible to find a pair of integers $(x,y)$ such that $f(x)=f(y)$ for $x\neq y$.
Hence, the legitimate user cannot detect the attacker's illegal acts.

Here,
we pay attention to the following fact.
When we specify the problem at first,
we assume that the function $f$ can be either one-to-one or two-to-one.
However,
if the legitimate user knows $f$ to be two-to-one beforehand,
he cannot obtain $s=0$ as an answer and he can conclude that it is evidence of attacker's illegal acts.

In Simon's algorithm,
the legitimate user obtains a set of random integers $\{k_{1},...,k_{m}\}$
from the measurements for the one-to-one function $f$ even if the attacker does not perform malicious acts.
Here,
we make the following assumptions.
If the legitimate user obtains $s\neq 0$,
he keeps $\{k_{1},...,k_{m}\}$ in a safe place carefully.
By contrast,
if he obtains $s=0$ and notices that $\{k_{1},...,k_{m}\}$ is a set of random integers,
he regards it as garbage data,
treat it carelessly,
and does not keep it secret.
In case the attacker disturbs the process of the algorithm and the legitimate user obtains a set of random integers $\{k_{1},...,k_{m}\}$,
we also assume that it is regarded as garbage data and not kept in a safe place.

Then,
the accomplice tells $\{k_{1},...,k_{m}\}$ to the attacker secretly behind a legitimate users's back.
The attacker eventually obtains correct $s$ from Eq.~(\ref{equations-for-s-attacker})
for two-to-one $f$ because he obtains $\{k_{1}\oplus j_{1},...,k_{m}\oplus j_{m}\}$.
Because
$\{k_{1},...,k_{m}\}$ is a set of random integers for the accomplice
during this sequence of illegal acts,
he cannot obtain any information about the exact solution $s$ at all.
Hence,
the attacker succeeds in cracking Simon's algorithm.

At the close of this section,
we consider a method of detecting the attacker's illegal acts.
If the legitimate user secretly runs an instance with a known two-to-one function $f$ and nontrivial $s(\neq 0)$,
he can promptly notice the attacker's malicious acts because of the random integers $\{k_{1},...,k_{m}\}$
observed on the quantum register A and their conclusion $s=0$.
We can apply a similar detection method to Shor's factoring algorithm.
It is discussed in Sec.~\ref{section-attack-Shor}.

\section{\label{section-attack-Shor}How to attack Shor's factoring algorithm}
In this section,
we discuss how to attack Shor's factoring algorithm.
First of all,
we assume that the composite number $N$ and the randomly chosen number $y$ in Eq.~(\ref{y-a-mod-N})
are publicly disclosed,
and the attacker knows them.
As explained in Sec.~\ref{section-attack-Simon},
the attacker replaces the qubits of the register A with the entangled ones that keep entanglement between quantum registers A and C.
We assume that the attacker holds the quantum register C.
Replacing the qubits of the quantum register A with entangled ones,
the attacker prepares the following state instead of Eq.~(\ref{Shor-1st-step}):
\begin{equation}
\frac{1}{\sqrt{q}}
\sum_{a=0}^{q-1}
|a\rangle_{\mbox{\scriptsize A}}
|a\rangle_{\mbox{\scriptsize C}}
|0\rangle_{\mbox{\scriptsize B}}.
\end{equation}
The legitimate user computes $F_{N}(a)$ given by Eq.~(\ref{y-a-mod-N}) and writes down its output on the quantum register B as
\begin{equation}
\frac{1}{\sqrt{q}}
\sum_{a=0}^{q-1}
|a\rangle_{\mbox{\scriptsize A}}
|a\rangle_{\mbox{\scriptsize C}}
|y^{a} \bmod N\rangle_{\mbox{\scriptsize B}}.
\end{equation}

Next,
the legitimate user observes the quantum register B, so that the quantum registers A and C irreversibly reduce to the following state
instead of Eqs.~(\ref{third-steo-Shor-algorithm-01}), (\ref{definition-A}), (\ref{third-steo-Shor-algorithm-02}), and (\ref{definition-f(a)}):
\begin{eqnarray}
&&
\frac{1}{\sqrt{A+1}}
\sum_{j=0}^{A}
|jr+l\rangle_{\mbox{\scriptsize A}}
|jr+l\rangle_{\mbox{\scriptsize C}} \nonumber \\
&=&
\sum_{a=0}^{q-1}
f(a)
|a\rangle_{\mbox{\scriptsize A}}
|a\rangle_{\mbox{\scriptsize C}}.
\end{eqnarray}
Then,
the legitimate user and the attacker apply the discrete Fourier transform to the quantum registers A and C,
respectively and independently,
and we obtain
\begin{equation}
\frac{1}{q}
\sum_{a=0}^{q-1}
\sum_{c=0}^{q-1}
\sum_{d=0}^{q-1}
\exp[2\pi ia(c+d)/q]f(a)
|c\rangle_{\mbox{\scriptsize A}}
|d\rangle_{\mbox{\scriptsize C}}.
\end{equation}

If the legitimate user observes the quantum register A,
he obtains $c$ such that
$0\leq c\leq q-1$
at random,
and he cannot detect the attacker's illegal acts.
The reason is as follows.
Even if the legitimate user performs Shor's algorithm properly without disturbance of the attacker,
a probability that he observes $c$ not satisfying Eq.~(\ref{range_rc_mod_q}) is equal to or less than $0.595$
as mentioned in Sec.~\ref{section-review-Shor}.

For the special situation where a period $r$ divides $q$ exactly,
the legitimate user observes $c$ satisfying Eq.~(\ref{range_rc_mod_q}) certainly.
Otherwise,
he observes $c$ not satisfying Eq.~(\ref{range_rc_mod_q}) with a non-zero probability.
We estimate this probability in Sec.~\ref{numerical-calculations}
numerically and analytically.
We find that it is approximately equal to $0.2263$.

Because of the above reason,
even if the legitimate user obtains a wrong period $r$,
he cannot insist that the attacker commits a crime.
Now,
we assume the following.
If the legitimate user obtains a wrong period $r$,
he regards $c$,
the result of the measurement of the quantum register A,
as garbage data and does not keep it in a safe place carefully.
Then,
we assume that the accomplice tells $c$ to the attacker secretly behind a legitimate user's back.
From a measurement of the quantum register C,
the attacker obtains $c+d$,
and he may eventually obtain a correct period $r$.

Here,
we examine how to detect the attacker's illegal acts.
The legitimate user can secretly run an instance with a known period $r$.
For example,
if the legitimate user tries to solve a problem where $r$ divides $q$ exactly,
he has to observe
$c=\lambda q/r$ for $\lambda=0,1,...,r-1$
for the quantum register A
with a zero error probability.
Thus,
he promptly notices the attacker's malicious acts.
The legitimate user can compute $a \bmod r$ instead of $F_{N}(a)$ given by Eq.~(\ref{y-a-mod-N}).
Clearly,
the function $a \bmod r$ has a period $r$.
If the legitimate user performs this detection method with a certain frequency,
he can notice the attacker's interference.
We pay attention to the fact that to prepare $F_{N}(a)$ with the known period $r$ is very difficult
because of a huge amount of calculations.

At the close of this section,
we consider the following.
The legitimate user discloses the composite number $N$ and the randomly chosen number $y$ in Eq.~(\ref{y-a-mod-N}).
Thus,
someone may insist that the attacker had better run Shor's algorithm by himself instead of attacking the legitimate user's quantum computer.
However,
cracking other person's quantum computer,
the attacker can execute Shor's algorithm without preparing a quantum circuit
which computes $F_{N}(a)$ defined in Eq.~(\ref{y-a-mod-N}).
In the process of the attack explained in this section,
the attacker only needs the entangling probe and the quantum Fourier transform.
This is merit for committing the illegal acts.

\section{\label{numerical-calculations}Numerical and analytical evaluations of the probability of obtaining $c$ that does not satisfy Eq.~(\ref{range_rc_mod_q})}
It is mentioned in Sec.~\ref{section-review-Shor} that the probability of obtaining $c$ that does not satisfy Eq.~(\ref{range_rc_mod_q})
for the measurement of the quantum register A is equal to or less than $0.595$.
In this section,
we evaluate the probability numerically and analytically.
Throughout this section,
we assume that the period $r$ does not divide $q$ exactly.

First, we estimate the probability numerically for concrete examples.
For example,
we consider $N=3\times 7=21$.
According to Sec.~\ref{section-review-Shor},
we take $L=9$ and $q=2^{L}=512$.
We choose $y=2$ for Eq.~(\ref{y-a-mod-N})
and obtain a period $r=6$.
Taking $l=0$ for Eq.~(\ref{third-steo-Shor-algorithm-01}),
we obtain $A=85$ for Eq.~(\ref{definition-A}).
The number of values of $c$ satisfying Eq.~(\ref{range_rc_mod_q}) is given by six and we define a set of them as
$S=\{0,85,171,256,341,427\}$.
The probability that we obtain $c$ with the measurement of the quantum register A can be written as $\mbox{Prob}(c)=|\tilde{f}(c)|^{2}$,
where $\tilde{f}(c)$ is given by Eq.~(\ref{Shor-amplitude}).
The probability of obtaining $c$ that does not satisfy Eq.~(\ref{range_rc_mod_q}) is given as follows:
\begin{equation}
P
=
1-\sum_{c\in S}\mbox{Prob}(c).
\label{Probability_exclude}
\end{equation}
From numerical calculations,
we obtain $P\simeq 0.2074$.

\begin{figure}
\begin{center}
\includegraphics[scale=1.0]{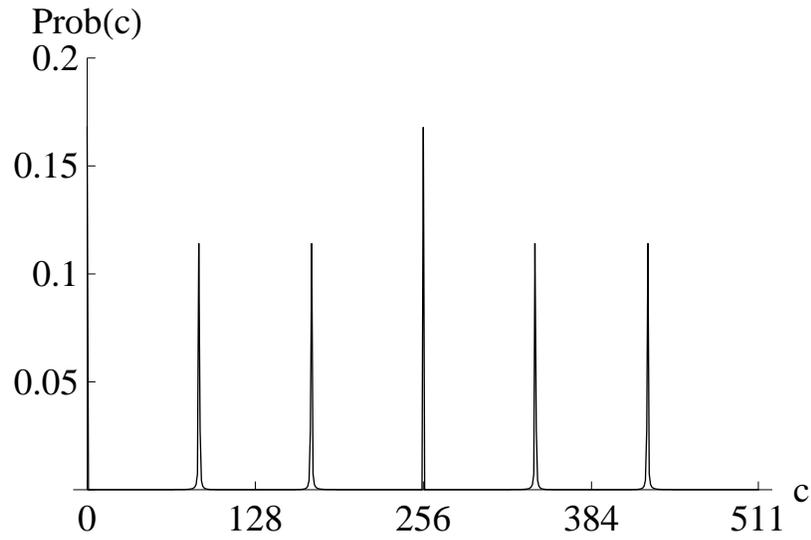}
\end{center}
\caption{A graph of probability distribution of $\mbox{Prob}(c)$ for $c\in\{0,1,...,511\}$
for $r=6$ and $l=0$.
A linear scale is used for the vertical axis.}
\label{Figure01}
\end{figure}

\begin{figure}
\begin{center}
\includegraphics[scale=1.0]{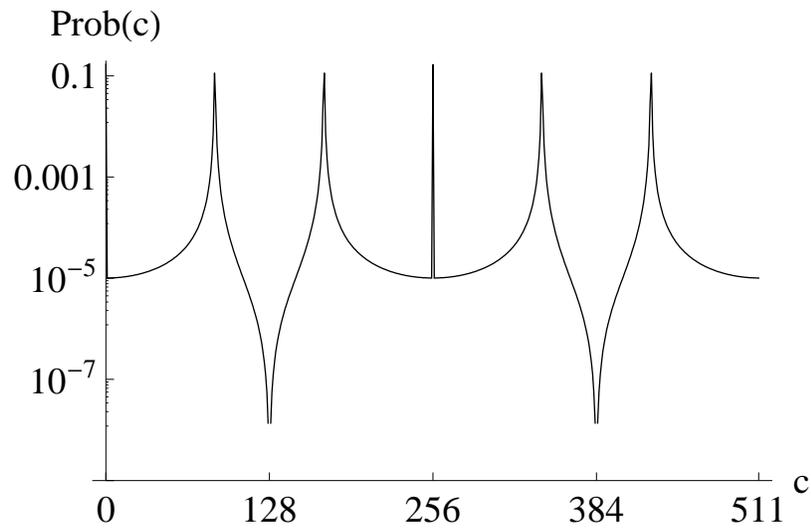}
\end{center}
\caption{A graph of probability distribution of $\mbox{Prob}(c)$ for $c\in\{0,1,...,511\}$
for $r=6$ and $l=0$.
A logarithmic scale is used for the vertical axis.
At many values of $c$,
$\mbox{Prob}(c)$ is larger than $10^{-5}$ around.}
\label{Figure02}
\end{figure}

We show probability distribution of
$\mbox{Prob}(c)$ for $c\in\{0,1,...,511\}$
for $r=6$ and $l=0$
in Figs.~\ref{Figure01} and \ref{Figure02}.
In Figs.~\ref{Figure01} and \ref{Figure02},
$\mbox{Prob}(c)$ is represented in linear and logarithmic scales, respectively.
Looking at Fig.~\ref{Figure02},
we notice that $\mbox{Prob}(c)$ is larger than $10^{-5}$ at many values of $c$.
Thus,
we can expect to observe random $c$ with a certain probability larger than $10^{-5}$ around.

\begin{table}
\caption{Numerical evaluations of $P$ defined in Eq.~(\ref{Probability_exclude}) for various composite numbers $N$.
We suppose $l=0$ for Eqs.~(\ref{third-steo-Shor-algorithm-01}) and (\ref{definition-A}) throughout all instances shown in this table.
The period $r$ for the function
$F_{N}(a)\equiv y^{a} \bmod N$ in Eq.~(\ref{y-a-mod-N})
is always a divisor of $\phi(N)$.
The probability $P$ approaches $P\simeq 0.2263$ as $N$ becomes larger.}
\label{Table01}
\begin{center}
\begin{tabular}{|c|c|c|c|c|c|c|}
\hline
$N$ & $L$ & $\phi(N)$ & $y$ & $r$ & $A$ & $P$ \\
\hline
$3\times 7=21$ & $9$ & $12$ & $2$ & $6$ & $85$ & $0.2074$ \\
$3\times 11=33$ & $11$ & $20$ & $2$ & $10$ & $204$ & $0.2204$ \\
$5\times 7=35$ & $11$ & $24$ & $2$ & $12$ & $170$ & $0.2099$ \\
$5\times 11=55$ & $12$ & $40$ & $2$ & $20$ & $204$ & $0.2204$ \\
$7\times 11=77$ & $13$ & $60$ & $2$ & $30$ & $273$ & $0.2243$ \\
$7\times 17=119$ & $14$ & $96$ & $3$ & $48$ & $341$ & $0.2099$ \\
$11\times 13=143$ & $15$ & $120$ & $2$ & $60$ & $546$ & $0.2251$ \\
$11\times 17=187$ & $16$ & $160$ & $3$ & $80$ & $819$ & $0.2204$ \\
$13\times 17=221$ & $16$ & $192$ & $3$ & $48$ & $1365$ & $0.2105$ \\
$13\times 19=247$ & $16$ & $216$ & $2$ & $36$ & $1820$ & $0.2245$ \\
$17\times 19=323$ & $17$ & $288$ & $3$ & $144$ & $910$ & $0.2243$ \\
$17\times 23=391$ & $18$ & $352$ & $3$ & $176$ & $1489$ & $0.2250$ \\
$19\times 23=437$ & $18$ & $396$ & $2$ & $198$ & $1323$ & $0.2263$ \\
$19\times 29=551$ & $19$ & $504$ & $2$ & $252$ & $2080$ & $0.2262$ \\
$23\times 29=667$ & $19$ & $616$ & $2$ & $308$ & $1702$ & $0.2261$ \\
$23\times 31=713$ & $19$ & $660$ & $3$ & $330$ & $1588$ & $0.2262$ \\
$29\times 31=899$ & $20$ & $840$ & $3$ & $420$ & $2496$ & $0.2262$ \\
$29\times 37=1073$ & $21$ & $1008$ & $2$ & $252$ & $8322$ & $0.2262$ \\
$31\times 37=1147$ & $21$ & $1080$ & $2$ & $180$ & $11650$ & $0.2262$ \\
$31\times 41=1271$ & $21$ & $1200$ & $3$ & $120$ & $17476$ & $0.2257$ \\
$37\times 41=1517$ & $22$ & $1440$ & $2$ & $180$ & $23301$ & $0.2262$ \\
$37\times 41=1517$ & $22$ & $1440$ & $16$ & $45$ & $93206$ & $0.2262$ \\
$37\times 43=1591$ & $22$ & $1512$ & $2$ & $252$ & $16644$ & $0.2262$ \\
$37\times 43=1591$ & $22$ & $1512$ & $9$ & $63$ & $66576$ & $0.2263$ \\
$41\times 43=1763$ & $22$ & $1680$ & $5$ & $420$ & $9986$ & $0.2263$ \\
$41\times 43=1763$ & $22$ & $1680$ & $10$ & $105$ & $39945$ & $0.2263$ \\
$41\times 47=1927$ & $22$ & $1840$ & $6$ & $920$ & $4559$ & $0.2262$ \\
$41\times 47=1927$ & $22$ & $1840$ & $16$ & $115$ & $36472$ & $0.2263$ \\
$43\times 47=2021$ & $22$ & $1932$ & $3$ & $966$ & $4341$ & $0.2263$ \\
$43\times 47=2021$ & $22$ & $1932$ & $4$ & $161$ & $26051$ & $0.2263$ \\
\hline
\end{tabular}
\end{center}
\end{table}

In Table~\ref{Table01},
we show numerical evaluations of $P$ defined in Eq.~(\ref{Probability_exclude})
for various composite numbers $N$.
Turning our eyes to Table~\ref{Table01},
we notice that the probability of Eq.~(\ref{Probability_exclude}) approaches
$P\simeq 0.2263$
as $N$ becomes larger.
This fact implies that we can find an exact period $r$ with a probability of $1-P\simeq 0.7737$
in Shor's algorithm.
The average number of trials for obtaining the exact $r$ is around
$\bar{n}=1/(1-P)\simeq 1.292$.
Because its standard deviation as the Poisson distribution is given by $\sqrt{\bar{n}}\simeq 1.137$,
the attacker can try illegal operation $1.137$ times per $\bar{n}\simeq 1.292$ proper operations.
Thus,
the expectation number of total trials for one illegal operation is given by
$(\bar{n}+\sqrt{\bar{n}})/\sqrt{\bar{n}}\simeq 2.136$.
Hence,
the attacker can perform the illegal operation once every $2.136$ operations at most or with a probability less than
$\sqrt{\bar{n}}/(\bar{n}+\sqrt{\bar{n}})\simeq 0.4681$.

Second,
we estimate the probability analytically.
Because we take $N\gg 1$ and $N^{2}\leq q\leq 2N^{2}$ and $r<N$ holds,
we can assume $r\ll q$.
Thus,
we can suppose $A+1\simeq q/r$ with neglecting a small roundoff error.
From Eqs.~(\ref{definition-f(a)}), (\ref{DFT-state-Shor}), and (\ref{Shor-amplitude})
and $\mbox{Prob}(c)=|\tilde{f}(c)|^{2}$,
we obtain
\begin{eqnarray}
\mbox{Prob}(c)
&\simeq&
\frac{r}{q^{2}}
|
\sum_{j=0}^{q/r-1}
\exp(ij\theta_{c})
|^{2} \nonumber \\
&=&
\frac{r}{q^{2}}
\left|
\frac{1-\exp(iq\theta_{c}/r)}{1-\exp(i\theta_{c})}
\right|^{2} \nonumber \\
&=&
\frac{r}{q^{2}}
\frac{\sin^{2}[q\theta_{c}/(2r)]}{\sin^{2}(\theta_{c}/2)},
\label{Prob-c-approximation}
\end{eqnarray}
where
\begin{equation}
\theta_{c}
=
\frac{2\pi}{q}(rc \bmod q).
\end{equation}
In the derivation of Eq.~(\ref{Prob-c-approximation}),
we neglect a factor $\exp(2\pi ilc/q)$
because it does not depend on $j$ and its absolute value is equal to unity.

We let $x=rc \bmod q$.
Then,
because of Eq.~(\ref{range_rc_mod_q}) and its related explanation mentioned in Sec.~\ref{section-review-Shor},
the number of values of $x$ that exist in an interval $[-r/2,r/2]$ is equal to $r$.
Here,
we assume that these values of $x$ are distributed in the interval $[-r/2,r/2]$ uniformly
and we can replace the summation $\sum_{c\in S}$ with an integral $\int_{-r/2}^{r/2}dx$
in Eq.~(\ref{Probability_exclude}).
Thus, we obtain
\begin{equation}
P
\simeq
1
-
2
\int_{0}^{r/2}
\frac{r}{q^{2}}
\frac{\sin^{2}(\pi x/r)}{\sin^{2}(\pi x/q)}dx.
\label{Probability_exclude-2}
\end{equation}
[In the case where $r$ divides $q$ exactly,
$rc \bmod q=0$ always holds for $c$ satisfying Eq.~(\ref{range_rc_mod_q}),
so that the values of $x$ are not distributed in the interval $[-r/2,r/2]$ uniformly.]

Because $0\leq x\leq r/2$, $r\ll q$, and $\pi x/q\ll 1$,
we have an approximation
$\sin^{2}(\pi x/q)\simeq (\pi x/q)^{2}$.
Then,
we can calculate $P$ in Eq.~(\ref{Probability_exclude-2}) as follows:
\begin{eqnarray}
P
&\simeq&
1
-
\frac{2r}{q^{2}}
\int_{0}^{r/2}
\frac{\sin^{2}(\pi x/r)}{(\pi x/q)^{2}}dx \nonumber \\
&=&
1
-
\frac{2}{\pi^{2}}
[-2+\pi \mbox{Si}(\pi)] \nonumber \\
&\simeq&
0.226{\,}305,
\end{eqnarray}
where
\begin{equation}
\mbox{Si}(x)
=
\int_{0}^{x}
\frac{\sin t}{t}dt.
\end{equation}
This analytical result corresponds well to the numerical one obtained above.

\section{\label{section-discussion}Discussion}
In this paper,
we examine how to attack Simon's algorithm and Shor's factoring algorithm
with entangling probes.
The attacks are performed in two stages.
At the first stage,
the attacker replaces the qubits of the quantum register with entangled ones.
At the second stage,
the accomplice tells results of measurements of the quantum register as classical information to the attacker secretly.
To succeed in cracking the quantum algorithms,
the attacker needs both quantum entanglement and classical information.
The situation looks like that of the quantum teleportation
\cite{Bennett1993}.
In the scenario of the quantum teleportation,
Alice and Bob share the maximally entangled state first and Alice sends a result of a measurement of qubits to Bob as classical information second.

In Secs.~\ref{section-attack-Simon} and \ref{section-attack-Shor},
we discuss methods of detecting the attacker's disturbance.
Because the attacker can commit a crime once every $2.136$ operations at most or with a probability less than $0.4681$
for Shor's algorithm,
the legitimate user has to carry out the detection test once every two operations around.
This is a cumbersome procedure for the legitimate user.

To protect the quantum computer against the attacks proposed in this paper,
we have to take care of the following three points.
First, we have to keep the initial state of the quantum register intact and sound.
We do not let it be entangled with a quantum system outside the institute.
Second, we have to keep the garbage data secret and erase it completely.
Third, we have to exclude the accomplice in the cracking from the institute.
Initializing the memory certainly,
treating the garbage data properly,
and guarding the system from an inside job.
These are key components of computer security in general.
From a viewpoint of the prevention of security incidents,
there is no difference between the classical and quantum computers.

\end{document}